\documentclass[10pt,journal,compsoc]{IEEEtran}
\IEEEoverridecommandlockouts

\usepackage[ right=0.85in, left=0.8in, top=1in, bottom = 1.0in, ]{geometry}
\usepackage{cite}
\usepackage{amsmath,amssymb,amsfonts}
\usepackage{algorithmic}
\usepackage{amsmath,amssymb,amsfonts}
\usepackage[linesnumbered,ruled,vlined]{algorithm2e}
\usepackage{algorithmic}
\usepackage{textcomp}
\usepackage{array}
\usepackage{flushend}
\usepackage{balance}
\usepackage{xcolor}
\usepackage{textcomp}
\usepackage{graphicx}

\SetKwInput{KwInput}{Input}                
\SetKwInput{KwOutput}{Output}              

\title{Self-Sovereign Identity for Trust and Interoperability in the Metaverse}

 \author{
\IEEEauthorblockN{
\textbf{Siem Ghirmai}, 
\textbf{Daniel Mebrahtom},
\textbf{Moayad Aloqaily}, \textit{Senior Member, IEEE}, 
~\textbf{Mohsen Guizani},~\IEEEmembership{Fellow, IEEE},~\textbf{Merouane Debbah},~\IEEEmembership{Fellow, IEEE}%
}
\IEEEcompsocitemizethanks{

\IEEEcompsocthanksitem \textit{S. Hadish, D. Mebrahtom, M. Aloqaily, M. Guizani} are with the Mohamed Bin Zayed University of Artificial Intelligence (MBZUAI), UAE. \protect E-mails: \{siem.hadish; daniel.gebre; moayad.aloqaily;\}@mbzuai.ac.ae, mguizani@ieee.org 
\IEEEcompsocthanksitem \textit{M. Debbah} is with the Technology Innovation Institute, Abu Dhabi, UAE. \protect E-mail: Merouane.Debbah@tii.ae
}}

\newcommand{\remove}[1]{}

\begin{document}

\maketitle
\begin{abstract}
  With the advancement in computing power and speed, the Internet is being transformed from screen-based information to immersive and extremely low latency communication environments in web 3.0 and the Metaverse. With the emergence of the Metaverse technology, more stringent demands are required in terms of connectivity such as secure access and data privacy. Future technologies such as 6G, Blockchain, and Artificial Intelligence (AI) can mitigate some of these challenges. The Metaverse is now on the verge where security and privacy concerns are crucial for the successful adaptation of such disruptive technology. The Metaverse and web 3.0 are to be decentralized, anonymous, and interoperable. Metaverse is the virtual world of Digital Twins and non-fungible tokens (NFTs).The control and possession of users’ data on centralized servers are the cause of numerous security and privacy concerns.This paper proposes a solution for the security and interoperability challenges using Self-Sovereign Identity (SSI) integrated with blockchain. The philosophy of Self-Sovereign Identity, where the users are the only holders and owners of their identity, comes in handy to solve the questions of decentralization, trust, and interoperability in the Metaverse. This work also discusses the vision of a single, open standard, trustworthy, and interoperable Metaverse with initial design and implementation of SSI concepts.
\end{abstract}


\section{Introduction}

Although it has been discussed for quite some time in one form or another, the Metaverse remains a vague concept for many people. Recently, Facebook has re-branded itself into Meta\cite{Facebook}, while a famous video game company - Epic Games announced a 1.0 billion dollars investment to build the Metaverse \cite{Epic-Games}, and other giant tech companies such as Google and Microsoft are in the race as well. When it comes to government investments, the Dubai government has also invested in the virtual world asset and is expected to start providing services within this virtual world.

Essentially, the "Metaverse" term generally refers to a universe beyond the physical world. Experts envision it as a 3D representation of the Internet which participants can access via Extended Reality (XR) technologies\cite{Encyclopedia}. Unlike Virtual Reality (VR) and Augmented Reality (AR) environments, which are completely separate artificial environments, the Metaverse is a fully immersive three-dimensional virtual world \cite{Immersiveness} where avatars engage in educational, economic, social, and cultural activities \cite{Avatars_activity}. It is built on technologies that allow for multisensory interactions with virtual worlds, digital objects, and people \cite{Encyclopedia}. Continuity of identity, shared environments, embodied avatars, synchronization, virtuality, interoperability, and immersive user experience are among the common attributes of the Metaverse \cite{Metaverse_attributes}. In the Metaverse, avatars with customizable appearances and behavior are used to represent people \cite{Avatar_behavior}. The Metaverse offers a wide range of activities to users such as playing virtual games, information exchange, meetings, socializing, and monetizing assets through NFTs \cite{Facebook}\cite{Avatars_activity}.

Despite the promising future of the Metaverse, a few issues that can limit the seamless integration of physical and virtual worlds in the Metaverse are yet to be fully addressed. The first issue is the need for establishing trust between users, as avatars can mimic the behaviors and features of other avatars, and people may not behave as expected \cite{Avatar_behavior}. To achieve the desired outcomes, Trust is regarded as a critical success factor in the Metaverse. The essential features of the Metaverse, such as interoperability, decentralization, immersiveness, and scalability, might indeed pose a variety of difficulties for trustworthy system provision \cite{Rich_content}. 

The second issue is the Interoperability of users' identities, data, and avatars between distinct virtual worlds in the Metaverse. As shown in Figure \ref{Metaverse Enablers}, the Metaverse's interoperability indicates the capacity to smoothly visit different virtual worlds in the Metaverse and move their data and assets to their preferred locations or Virtual Service Provider (VSP). A non-interoperable set of virtual worlds might limit the user’s identity, avatars, and data to a specific VSP. This would be undesirable and inconsistent with the vision for Web 3.0 and the Metaverse. 

To tackle those issues, this paper explores the idea of Self-Sovereign Identity (SSI) and explains ways where this technology can be implemented to enable interoperability between Virtual Worlds (VWs). It also discusses ways to ensure the trust between different entities involved in the Metaverse relying on a blockchain-based SSI management system.

The contributions of this paper are summarized as follows:

\begin{itemize}
  \item We discuss the requirements, enabling technologies, and related standards for implementing the Metaverse.
  \item We discuss the role of Digital Twin (DT), Blockchain, 6G, Extended Reality (XR), machine learning (ML), and wearable sensors in implementing a user-centric Metaverse successfully.

\item We propose the Metaverse as a single, open standard that provides a super-immersive 3D environment supporting seamless integration of the physical and virtual worlds in the 6G era.
 \item Then, we motivate the need for trust between different stakeholders.
 \item Finally, we present a design and an initial implementation using SSI to solve the concerns of decentralization, trust, and interoperability in the Metaverse by giving users full ownership, control, and possession of their digital identity and data.
\end{itemize}

The remainder of this paper is organized as follows. Section \ref{rel} investigates the state-of-the-art. Section \ref{Overview} presents an overview of the Metaverse implementation requirements, enabling technologies, and metaverse-related existing standards. The significance of the SSI from different aspects is discussed in section \ref{SSI}. Then, we discuss the proposed approach in Section \ref{Proposed}. Finally, we draw the conclusion and future directions in Section \ref{conclusion}.

\section{Related Work} \label{rel}
Despite the significance of a real-time interoperable and decentralized Metaverse, only limited research has been conducted on this topic. To begin with, the authors in \cite{Immersiveness} discuss the idea of interoperability in the virtual world and the Metaverse, and its important role in enabling and facilitating information exchange or interacting with each other seamlessly and transparently. They regard interoperability as enabling technology that permits users to benefit from the seamless transfer of identities from one point (one sub-Metaverse) to another with no interruption of experience, similar to how we physically move between distinct physical locations in the real world. With interoperability, therefore, users may move around with full access to any environment “without the disruption of changing login credentials or losing one’s chain of cross-cutting digital assets.” They furthermore emphasize a set of standards associated with the interoperability of various layers in the virtual world, such as model standards, protocol standards, locator standards, identity standards, and currency standards. 

In \cite{blockchain_for_Meta}, the authors discuss the idea of integrating blockchain in the Metaverse aiming to achieve interoperability. They argue that interoperability will be among the main driving forces behind the Metaverse, and blockchain is expected to make it possible for exchanging data located in different sub-Metaverse. For example, finance and healthcare virtual environments will be able to communicate and exchange data. This way, “users will be able to keep their avatars and possessions while easily transferring them between virtual worlds” \cite{blockchain_for_Meta} (p.8). They highlight the current interoperability challenges in the Metaverse raised due to the fact that the current digital realms employed traditional, centralized, disjointed, and unorganized platforms. People are required to create accounts, avatars, and wallets to participate in these realms. They suggested that cross-chain protocol can be employed to achieve interoperability between virtual worlds, therefore; individuals can exchange possessions like avatars, NFTs, and currencies between virtual worlds. 

Le et al. \cite{interoperability} discuss the significance of interoperability in the Metaverse for gaming purposes. Thus, interoperability is a key requirement for the Metaverse for supporting the creation and distribution of content through distinct virtual environments. For example, content created in Minecraft can be transferred to another gaming environment (e.g. Roblox) with continued identity and experience frictionlessly.  The Metaverse is perceived as the digital twin of the real world, interoperability enables users to keep their avatars’ attributes when accessing distinct virtual environments in the Metaverse, and offers more freedom to the user.  With that being said, the authors call on organizations to build protocols and standards that provide common grounds to connect distinct virtual environments in the Metaverse, like the TCP/IP for the Internet. Furthermore, the Open Metaverse Interoperability Group \cite{Group}, which works to establish technological standards, is constantly working to bridge virtual worlds by designing and promoting protocols for identity, social graphs, inventory, and more. 

Table \ref{summary of related work} summarizes the literature from six aspects: 1) Significance 2) Technical requirements for developing a Metaverse 3) Interoperability, 4) Metaverse related standards 5) Trust , and 6) Security and Privacy in the Metaverse?

\begin{table*}[htbp]
\centering
\caption{Summarized the comparison between the current related work and this paper}
\label{summary of related work}
\begin{tabular}{p{1.0in}|p{1.9in}|p{0.5in}|p{0.6in}|p{0.5in} |p{0.7in}|p{0.3in}}\\
    \hline
    \textbf{Article} & \textbf{Contribution} & \textbf{Required Tech?} & \textbf{Proposed Model?} & \textbf{Standards?}  & \textbf{Security and Privacy?} & \textbf{Trust?}\\
    \hline
     Dionisio \textit{et al.} \cite{Immersiveness}  & Focused on immersive realism, the ubiquity of access and identity, interoperability, and scalability of Metaverse & Yes & No & Yes & No & No\\
    \hline
    Gadekallu \textit{et al.} \cite{blockchain_for_Meta}    & Investigate the role of blockchain in the Metaverse from a technical perspective, challenges and solutions & Yes & No  & No & Yes &Yes \\
    \hline
    Lee \textit{et al.} \cite{interoperability} & Discuss user-centric factors for the development of the Metaverse & Yes & No & No & Yes & Yes  \\
    \hline
    Open Interoperability Group\cite{Group} & Develop a common protocol to connect individual virtual spaces & No & No & Yes & No & No \\
    \hline
    \textbf{Proposed Model} & Proposed a model that ensures Trust and Interoperability in the Metaverse  & Yes & Yes & Yes & Yes & Yes \\
 \hline 
\end{tabular}
\end{table*}

\section {Metaverse implementation: Overview} \label{Overview}
Although the idea of Metaverse is in its early stage, several VWs implementations, including Second Life, Meta Horizons, Fortnite, Decentraland, Nvidia Omniverse, Roblox, Otherside, and The Sandbox, are being developed in a non-interoperable manner with their avatars, ecosystems, and currencies by cooperating the several technologies \cite{Metaverse_attributes}. The future Metaverse is envisioned as a single entirely interconnected 3D decentralized network in which all the sub-Metaverse co-exist together in a way that allows users to seamlessly move from one part to another, providing the best possible experience for users who want to move through the virtual world using their avatar, with no one owning it \cite{Meta_ownership}.

We dedicate this section to introducing the Metaverse from four different aspects:  general requirements,  current enabling technologies, and Metaverse-related standards.

\subsection{Requirements}
There is a set of requirements that one should consider when building a reliable, interactive, scalable, secure, and interoperable Metaverse. Ideally, the Metaverse should have:
\begin{itemize}
 \item \emph{Responsive Vision and Ergonomics}: The Metaverse should pack a high-resolution headset with low latency time to avoid motion sickness.
 \item \emph{ Open Interoperability}: Users should be able to use their avatars, digital wallet, and objects across multiple virtual worlds.
 \item \emph{Open Standard}: Experts envision it as a 3D representation of the internet in which participants can access extended reality (XR) technologies. Like the Internet, no single entity should take control of the Metaverse.

\item \emph{Security and Privacy}: Like online social networks, avatar impersonation, user authentication \cite{girmay2021ai}, online harassment, NFTs security, and copyright issues are expected to occur in the Metaverse as well. Standards and regulations should be adopted to govern data collection for a personalized immersive experience in the Metaverse.     
\item  \emph{Immersive Experience}: The entire human body should be able to feel a multi-directional movement, along with immersive audio for enhanced and realistic user experience.
 
 \item \emph{Decentralized Economy}: Freedom of fair trade of assets, cryptocurrency, and NFTs should be granted to users for the creation of a decentralized economy.  
\end{itemize}

\subsection{Current Enabling Technologies}
The Metaverse employed several cutting-edge technologies and enablers for creating an interactive 3D virtual user experience.

\begin{itemize}
    \item \emph{Reliable 6G Network}: A huge amount of sensing data is generated and transmitted to be further processed between the real world and the virtual world in the Metaverse. Tang et al.\cite{6G} emphasized that the 6G network has the bandwidth and capacity to support ultra-high communications. To establish real-time information exchange in the Metaverse, the 6G has low ultra-latency in terms of communication and processing speed. 
    \item \emph{Digital Twin}: It is a digital representation that represents the real world with high fidelity and accuracy\cite{alo}. In short, it is the digital replica of a real-world physical object. Using digital twins, it is possible to analyze, predict, and optimize the potential outcomes of real-world physical entities in advance through simulation of situations that might occur in real life.
    
    \item \emph{Extended Reality (XR)}: From a technical point of view, Virtual Reality (VR), Augmented Reality (AR), and Mixed Reality (MR) fall under the XR umbrella. In addition to offering an immersively simulated experience without physical limitations, VR enables the user to act as avatars in a digitally generated 3D world \cite{Avatar_behavior}. While the VR is a completely digital environment, the AR overlays real-world object settings into virtually created sounds, images, graphics, 3D models, videos, games, and GPS information. The MR is a concept that integrates the the VR and the AR to blend the real and virtual worlds, providing an enhanced, interactive, and more realistic user experience by allowing the user to interact with real-world objects while immersed in a virtual environment \cite{Avatars_activity}.

\begin{figure*}[htbp]
\centering
    \includegraphics[width=0.8\linewidth]{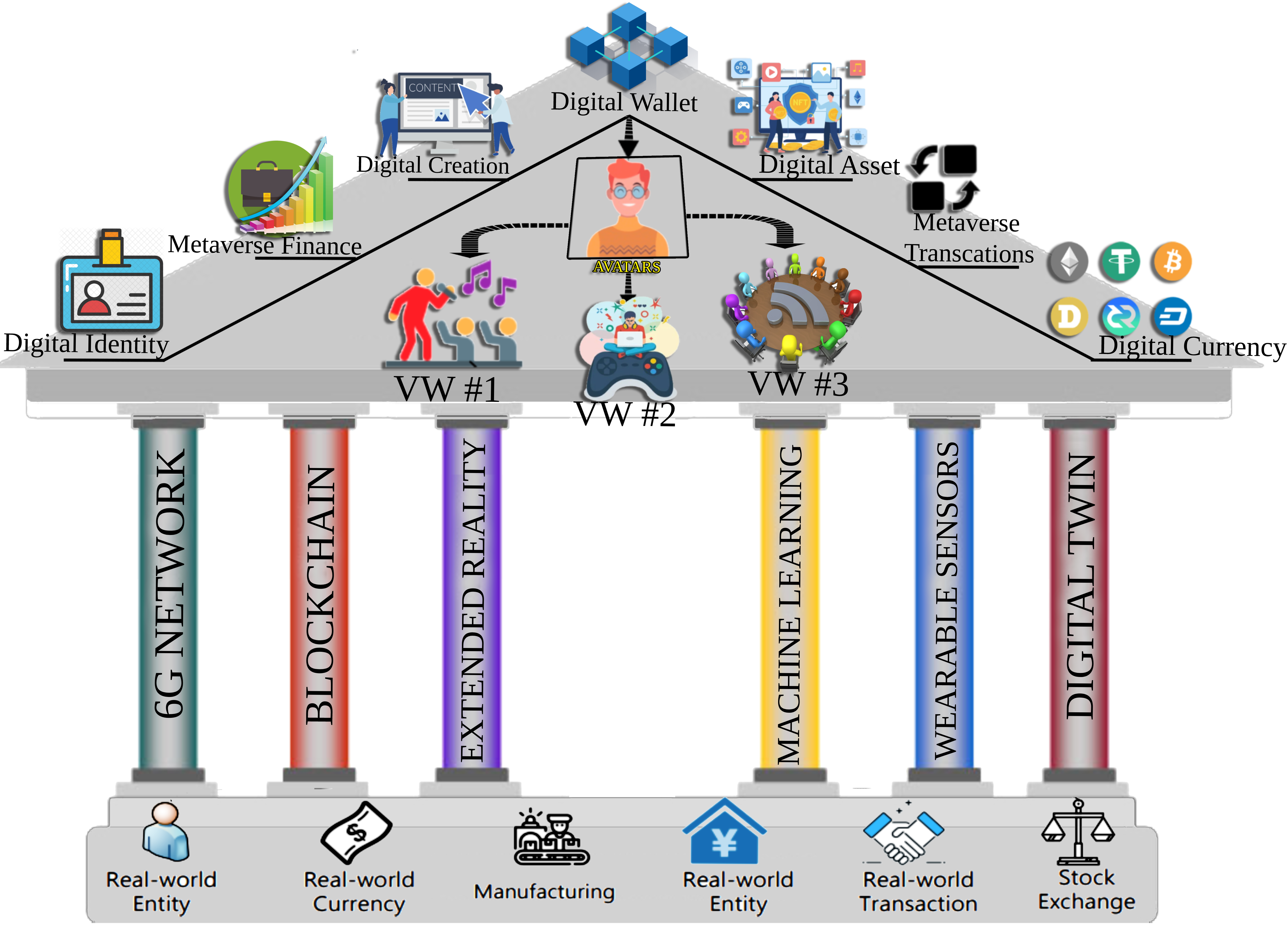}
    \caption{The architecture of a single, interoperable, and open standard Metaverse. The pillars demonstrate the enabling technologies for a successful deployment of the  Metaverse. The bottom layer represents the real world.}
    \label{Metaverse Enablers}
\end{figure*}
   
    \item \emph{Artificial Intelligence (AI)}: To make the most of the Metaverse, the AI is among the ultimate factors for extending and enhancing user cognitive space experience\cite{bou}. AI/ML models can be employed in the Metaverse for analyzing, recognizing, and predicting complex wearable sensors and other human-machine interaction gadgets' movements. Moreover, AI enables cognitive computer vision, content analysis, supervised speech processing, automatic resource allocation, attack prevention, sentiment analysis,  and 3D object rendering \cite{blockchain_for_Meta}.  
    \item \emph{Blockchain}: The blockchain system ensures the integrity and authenticity of the information in a decentralized fashion \cite{nakamoto}. In the context of the Metaverse, blockchain is one of the key enablers applied in different arrays of functionalities including but not limited to ensuring data privacy, ensuring data security, enabling data interoperability, ensuring data integrity, creating financial systems, smart contract deployment, handling a huge amount of data, digital proof of ownership, digital collectibility, value transfer, and governance and NFTs \cite{bc}. Blockchain would be central in storing and validating digital contents and enabling their trade to other platforms as opposed to the current e-commerce systems where their value is limited to centralized platforms \cite{Big_data}. Blockchain also assists in the trading of digital arts through NFTs \cite{Epic-Games}.
    \item \emph{Wearable Sensors}: One of the technically limiting barriers to immersiveness and realism in the Metaverse is the Hardware components. Specialized IoT sensors support multimodal immersion by providing sensory information such as sight, sound, touch, temperature, smell, balance, and gesture to the brain\cite{Encyclopedia} aiming to extend the ability of users to operate in and navigate through the Metaverse both physically and virtually \cite{blockchain_for_Meta}. Head-mounted display (HMD), goggles, sensor gloves, and other haptic devices are considered essential hardware components that enhance the sense of immersion in the Metaverse. 
\end{itemize}

\subsection{Metaverse Related Standards}
In prior studies, only two Metaverse-related standards existed: IEEE 2888 \cite{IEEE_2888} and ISO/IEC 23005  \cite{ISO_23005}.
\begin{enumerate}
\item \emph{IEEE 2888}:
This is a family of standards that defines the guidelines for bridging the Cyber and physical world. It has a set of standards such as: 
2888.1, 2888.2, 2888.3, 2888.4, 2888.5, and 2888.6. In general, these standards define the vocabulary, requirements, metrics, data formats, and APIs for acquiring information from sensors, enabling the definition of interfaces between the Cyber world and the physical world. The IEEE 2888 standard is expected to play a significant role in the building and success of the Metaverse. 

\item \emph{ISO/IEC 23005}:
This standard ensures interoperability between virtual worlds by providing architecture and specifying associated information representations to facilitate the interactions between the digital contents (gaming, simulation) and the real world (sensors, actuators, vision, and rendering). Several Metaverse-based services, such as audiovisual information and rendered sensory effects could benefit from this standard \cite{Big_data}.  
\end{enumerate}

\section {Self Sovereign Identity (SSI) Scheme} \label{SSI}
In the current Internet technology, identity management, data management, and authentication are done separately for different service providers mostly using emails and passwords. This Internet architecture where specified companies store all the information owned by the company and its users has numerous drawbacks related to the privacy and security of users' information. The companies using Web 2 technologies gather up and use huge amounts of data from users and to use their service, clients are expected to trust the central companies for their privacy. Single point of failure, violation of Zero trust, and the creation of monopolies, therefore, prompt us to look for ways to eliminate those issues in future technologies. SSI proposes a blockchain mechanism to replace traditional authentication that uses passwords with a decentralized authentication method that utilizes Zero Knowledge Proofs where the information used to authenticate the user is not transferred to the authenticator \cite{ssibac}. According to \cite{ssibac}, the SSI based access control system implements Verifiable Credentials, machine-readable credentials that are issued by one or more issuers, are possessed by a holder and are then used to derive the verifiable presentation which is presented to the verifier \cite{ferdous2019search}. This way, the authentication information does not need to be transferred through the Internet and stored on the servers of different companies. With the availability of their identity on their hand, users can then use the same information to communicate with different websites. In this work, we look for ways with similar mechanisms of sharing the same identity with different service providers. This can be improved to allow the owning of all the private data and use it to access different services in different virtual worlds. The questions this paper tries to answer are:
\begin{enumerate}
\item What are the benefits of implementing the SSI mechanism in the context of the decentralized Metaverse?

\item How would key functionalities like authentication and communication work in combination with SSI concepts in an interoperable Metaverse?
\end{enumerate}

\begin{figure*}[htbp]
\centering
    \includegraphics[width=0.8\linewidth]{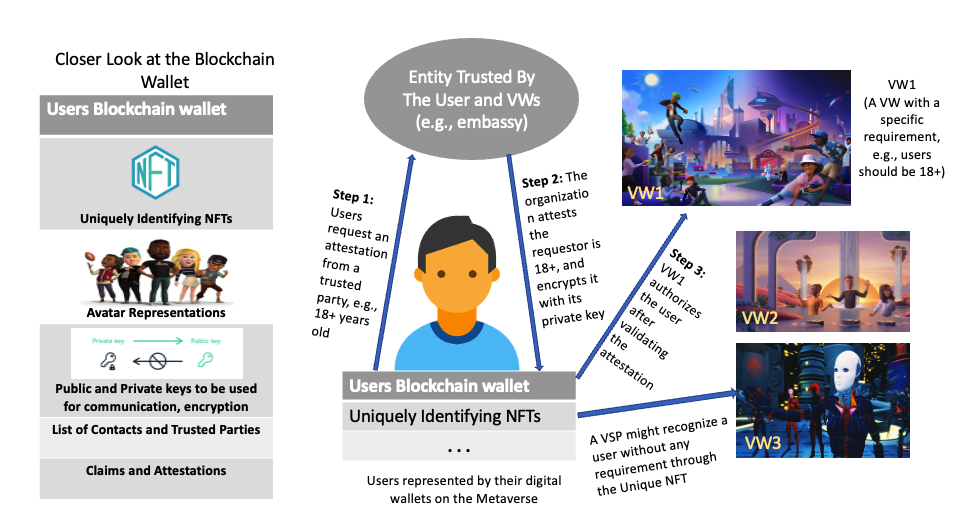}
    \caption{General description of SSI for the Metaverse in action. Two scenarios are considered and described for the authentication of a user.}
    \label{SSI for the Metaverse in action }
\end{figure*}

\section{Proposed Model} \label{Proposed}

The idea of the Metaverse having to be decentralized and interoperable fits well with the concept of the SSI where the users have their credentials and use them to be authenticated for different services and organizations without having to give their information. In this work, we discuss how all the information related to the user’s representation on the Metaverse can and should be owned and stored with the user as opposed to storing it with centralized organizations. In using the SSI scheme, a user can hold a verifiable claim and then uses cryptographically secure methods to prove it without giving out any information complying with the Zero Proof knowledge. On the Metaverse, all the information which is owned by the user can be owned and stored on the blockchain out of access of centralized organizations, and whenever a user needs a specific service, they can attest to the required data as their own and use cryptographic methods. This is similar to the Diffie-Heilman key exchange and is used to communicate with the agent that needs to know this information without giving up their data. For example, two users chatting on the Metaverse can attest their public keys to each other following the SSI principles and communicate a pre-shared key to encrypt their conversation. For the system suggested, the users will need to hold an NFT that is unique to their wallet, and which is untransferable to other wallets, a set of public and private keys to encrypt their communication with different organizations, and these would be the basis for implementing an interoperable Metaverse with different virtual worlds. Different VWs will require different levels of authentication and Trust between them and their clients. In the following section, we discuss the authentication aspect of the proposed system. 

\subsection {Authentication}
According to \cite{sawers}, an identity and an authentication mechanism can not be owned by the same organization as the authenticating organization ceasing to exist means that the identity would also stop existing. They propose that a user would hold a specific NFT that is immutable and nontransferable to other wallets, and this would be the unique identifier and would be hard to impersonate. A gallery that allows people to join it without any constraint would just need to recognize that authentication NFT,  and when the user is visiting for the second time, one would be recognized. This method of authentication however would not be enough when an extra level of trust is required. Another gallery, for example, would require its attendants to be above the age of 18, and in this case, a simple NFT will not provide enough evidence for the service provider to trust the requesting user. This technology combined with the concepts of SSI would, however, build up trust by having a mediator trusted by both entities attest to the claim of the requestor and issue the service provider enough evidence to trust the user as illustrated in Figure 2. A government agency for example might sign that a user is above the age of 18 and any organization might know that the person is of age, when the agency disappears from the Internet, the user would still have the assessment of his or her age. This type of authentication is not specific to an environment or organization and does not require the transfer of real-world knowledge to the VW service providers. 

\subsection {Avatar Recognition}

When people meet in the Metaverse, they would see the user’s avatars, the user’s real or computationally modified voices and movements. These attributes can easily be replicated by impersonators with the help of AI bots and used in the identity theft after the authentication step \cite{battle}. These impersonated avatars might not be identical to the original computationally. But the visual, auditory, and behavioral similarities are enough to trick other users. To counter this issue, a mechanism similar to the authentication done by the Metaverse environment can be done. When people first meet, they can share their identifying NFT and save each other's information similar to how contacts are saved on modern-day cell phones. If the meeting step requires an extra level of trust, for example, there is the buying and selling of digital assets, another mutual contact trusted by both users can attest to each other's authentication NFTs. From the first meeting onwards, the users have each other’s address, and from that point, the requirement is to visualize the user’s universally unique identifier, UUID, or the contact name or code saved by the other user when the two avatars see one another or are communicating in any way. 

\begin{figure*}[htbp]
\centering
    \includegraphics[width=0.8\linewidth]{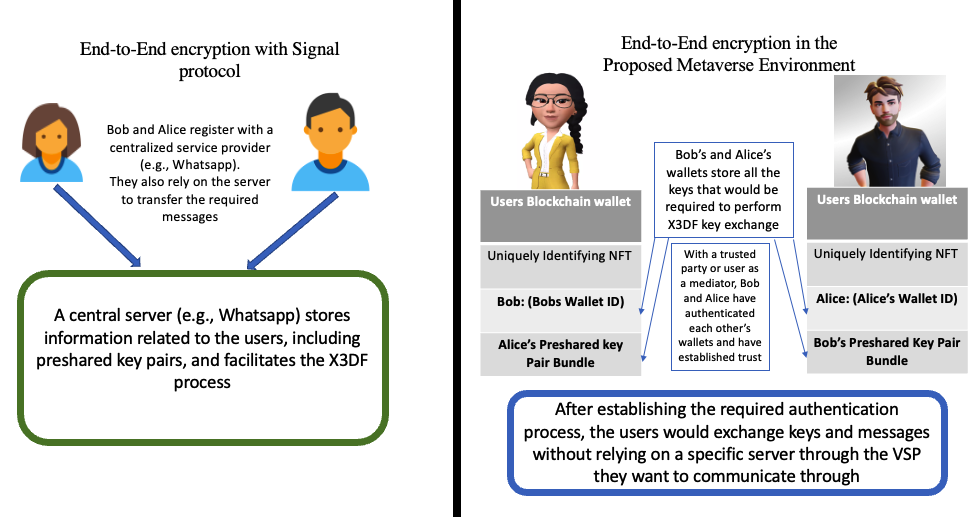}
    \caption{Comparison of End-to-End encryption using Signal protocol vs the proposed system}
    \label{comparison}
\centering
\end{figure*}

\subsection{Communication and End-to-End encryption}
In a decentralized and open Metaverse environment where the users would provide more data than ever before to the virtual service providers (VSPs), and where extra layers of technologies are being added to the communication channels adding more potential exploits, it is natural to be extra cautious with the communication channels being used. The current communication applications like WhatsApp use the Signal protocol where central servers manage the public and private keys required to encrypt and decrypt the messages \cite{signal}. In the Metaverse, however, the Signal protocol might not be as useful as the communication on the Metaverse should be decentralized and must be interoperable. As displayed in Figure 3, centralized servers such as in WhatsApp act as the mediator to authenticate and manage the public and private key pairs used to perform the Extended triple Diffie-Heilman key exchange (X3DH) and need to have both the communicating parties as clients to its server. With the help of the SSI however, provided that they have built trust between eachother, users can self-attest the public keys they want to use for communication with another user and perform the triple Diffie-Heilman key exchange used in the Signal protocol to generate a pre-shared key used to encrypt the messages. After a user requests a secure communication channel as described in Algorithm 1, the second user runs Algorithm 2 which might then be followed with other key derivation functions on both devices.







\begin{algorithm}[t]
\LinesNotNumbered
\caption{End-to-end Encryption with SSI (Requesting user)}
\label{alg:etee}
\KwInput{Receiver's Identity NFT $RID_{n_{f_t}}$}
\KwInput{Requester's Identity NFT $ReqID_{n_{f_t}}$}
\KwInput{ID of the Trusted Party $TP_{I_D}$, Attestation Certificate $Cert_{A_t}$}
\KwInput{Receiver's signed public keys bundle $SPK_{r_{e_c}} = spk_1, spk_2, ..., spk_N$}
\KwInput{Fetch the Receiver's public key $Rec_{p_k}$, Trusted Party's Public key $TP_{p_k}$, from the SSI blockchain $SSI_{B}$}
\If{$Cert_{A_t}$ is validated by $TP_{p_k}$}
{
    send $SPK_{r_{e_c}}$, $ReqID_{n_{f_t}}$, $TP_{I_D}$, and $Cert_{A_t}$ to the receiver's address\;
    \For{$spk_i$ in $SPK_{r_{e_c}}$}
    {
        \If{$TP_{I_D}$ validates $spk_i$}
        {
            \Return X3DH($spk_{i}$, $Req_{primkey}$, $Req_{pub}$)\;
        }
    }
}
\Else
{
    \Return None\;
}

\end{algorithm}






\begin{algorithm}[t]
\LinesNotNumbered
\caption{End-to-end Encryption with SSI (Replying User)}
\label{alg:etee2}
\KwInput{Requester's Identity NFT $RID_{n_{f_t}}$}
\KwInput{ID of the Trusted Party $TP_{I_D}$, Attestation Certificate $Cert_{A_t}$}
\KwInput{Requestor's signed public keys bundle $SPK_{req}= spk_1, spk_2,\dots,spk_N$}
\KwInput{Fetch the Requester's public key $Req_{p_k}$, Trusted Party's Public key $TP_{p_k}$, from the SSI blockchain $SSI_{B}$}
\If{ $Cert_{A_t}$ is validated by $ TP_{p_k}$ } {
    \For{$spk_i$ in $SPK_{req}$} {
        \If{$TP_{I_D}$ validates $spk_i$} {
            \Return X3DH($spk_i$, $U_{primkey}$, $U_{pub}$)
        }
    }
}
\Else {
    \Return None\;
}
\end{algorithm}

\section{Conclusion and Future Directions} \label{conclusion}

The concept of the Metaverse has existed for quite some time, with a predicted market value worth billions of dollars in the next few years. Despite the promising future of the Metaverse, a lack of trustworthiness and interoperability among the virtual world inside the Metaverse hinders its progress. Hence, in this paper, we present our vision of a single, open standard, trustworthy, and interoperable Metaverse, after a thorough investigation of the literature from three perspectives: General requirements, Enablers, and existing Metaverse-related standards. We then discuss how the proposed SSI scheme can be improved and employed in the Metaverse to achieve interoperability and trust. In the future, we plan to extend this work along the following research directions:

\begin{itemize}

\item In this paper, the objects discussed to be made interoperable are assumed to be data points to be read and updated. For example, when an avatar is mentioned to be interoperable, we mean the skin, clothing, and other similar attributes that might not necessarily have behaviors. Further studies can be made looking into making objects with behaviors interoperable between VWs.
\item The end-to-end encryption mentioned in the proposed model is based on live communications where both users are communicating directly, this might also be broadened to include communication when one of the users is offline, and investigate the encryption mechanisms when the users involved are more than two.
\item In addition, blockchain storage options might be discussed as the users controlling and managing their data and identity will require a different mechanism of storage and all the data might be too large to be stored on the blockchain.

\end{itemize}
We also aim to fulfill the idea of building a simulator to implement our proposed idea. The planned Metaverse should be built with a few VWs inside it, that support seamless interoperability between the VWs in a trustworthy manner. 


 
\bibliographystyle{IEEEtran}
\bibliography{References}

\end{document}